\input amstex
\documentstyle{amsppt}
%----------------------------------------------------------------
% Title:     On linear regression in three-dimensional Euclidean space.
% Authors:   O. V. Ageev, R. A. Sharipov
% Comments:  AmSTeX, 4 pages, amsppt style
% MSC-class: 51N20, 68W25
%----------------------------------------------------------------
%           Replacement for output macro definition
%
\catcode`@=11
\redefine\output@{%
  \def\break{\penalty-\@M}\let\par\endgraf
  \ifodd\pageno\global\hoffset=105pt\else\global\hoffset=8pt\fi  
  \shipout\vbox{%
    \ifplain@
      \let\makeheadline\relax \let\makefootline\relax
    \else
      \iffirstpage@ \global\firstpage@false
        \let\rightheadline\frheadline
        \let\leftheadline\flheadline
      \else
        \ifrunheads@ %\let\makefootline\relax
        \else \let\makeheadline\relax
        \fi
      \fi
    \fi
    \makeheadline \pagebody \makefootline}%
  \advancepageno \ifnum\outputpenalty>-\@MM\else\dosupereject\fi
}
\def\Beta{\mathchar"0\hexnumber@\rmfam 42}
\redefine\mm@{2010} % Math Subject Classification year
\catcode`\@=\active
%----------------------------------------------------------------
\nopagenumbers

\def\negskp{\hskip -2pt}
%----------------------------------------------------------------

%----------------------------------------------------------------
\def\blue#1{#1}
\catcode`#=11\def\diez{#}\catcode`#=6
\catcode`&=11\catcode`&=4
\catcode`_=11\def\podcherkivanie{_}\catcode`_=8
\catcode`\^=11\catcode`\^=7
\catcode`~=11\catcode`~=\active
\def\mycite#1{\cite{\blue{#1}}\immediate\special{ps:
     ShrHPSdict begin /ShrBORDERthickness 0 def}}

\def\mytag#1{%
    \tag#1}
\def\mythetag#1{\thetag{\blue{#1}}\immediate\special{ps:
     ShrHPSdict begin /ShrBORDERthickness 0 def}}
\def\myrefno#1{\no#1}
\def\myhref#1#2{\blue{#2}\immediate\special{ps:
     ShrHPSdict begin /ShrBORDERthickness 0 def}}

\def\mytheorem#1{\csname proclaim\endcsname{Theorem #1}}
\def\mytheoremwithtitle#1#2{\csname proclaim\endcsname{Theorem #1#2}}
\def\mythetheorem#1{\blue{#1}\immediate\special{ps:
     ShrHPSdict begin /ShrBORDERthickness 0 def}}
\def\mylemma#1{\csname proclaim\endcsname{Lemma #1}}
\def\mylemmawithtitle#1#2{\csname proclaim\endcsname{Lemma #1#2}}

\def\mycorollary#1{\csname proclaim\endcsname{Corollary #1}}

\def\mydefinition#1{\definition{Definition #1}}

\def\myconjecture#1{\csname proclaim\endcsname{Conjecture #1}}
\def\myconjecturewithtitle#1#2{\csname proclaim\endcsname{Conjecture #1#2}}

\def\myproblem#1{\csname proclaim\endcsname{Problem #1}}
\def\myproblemwithtitle#1#2{\csname proclaim\endcsname{Problem #1#2}}

%----------------------------------------------------------------
\pagewidth{360pt}
\pageheight{606pt}
\vphantom{a}
\vskip -0.5cm
\topmatter
\title
On linear regression in three-dimensional Euclidean space.
\endtitle
\author
O. V. Ageev, R. A. Sharipov
\endauthor
\address Self-employed individual, Ufa, Russia
\endaddress
\email ageev-ufa\@yandex.ru
\endemail

\address Bashkir State University, 32 Zaki Validi street, 450074 Ufa, Russia
\endaddress
\email r-sharipov\@mail.ru
\endemail
\urladdr
\vtop to 20pt{\hsize=280pt\noindent
\myhref{http://ruslan-sharipov.ucoz.com}
{http:/\negskp/ruslan-sharipov.ucoz.com}\newline
\myhref{http://freetextbooks.narod.ru}
{http:/\negskp/freetextbooks.narod.ru}\vss}
\endurladdr
\abstract
    The three-dimensional linear regression problem is a problem of finding 
a spacial straight line best fitting a group of points in three-dimensional 
Euclidean space. This problem is considered in the present paper and a solution 
to it is given in a coordinate-free form.
\vskip - 1cm
\endabstract
\endtopmatter
\TagsOnRight
\document

\rightheadtext{On linear regression \dots}
\head
1. Introduction.
\endhead
    The linear regression problem in two-dimensional case (i\.\,e\. on a plane) typically 
arises when approximating experimental data with a linear function (see \mycite{1}). 
Its solution using least squares method was first published by Legendre in 1805
(see \mycite{2}). In an unpublished form the least squares method is attributed
to Carl Friedrich Gauss 1795. His work was published only in 1809 
(see \mycite{3}).\par
     There are various fitting problems in three-dimensional Euclidean space (see plane,
circle and ellipse fitting problems in \mycite{4} and \mycite{5}, see ellipsoid fitting 
problem in \mycite{6} and \mycite{7}). The linear regression problem in three-dimensional 
case is the problem of best fitting some straight line to a group of points in three-dimensional Euclidean space. A solution of this problem is given by Jean Jacquelin
in \mycite{8}. His method is substantially based on direct calculations using coordinates.
Our goal in the present paper is to give a coordinate-free solution to the problem. 
\head
2. Parametric and non-parametric vectorial equations of a straight line.
\endhead
\parshape 3 0pt 360pt 0pt 360pt 180pt 180pt
     Let's consider the straight line $AX$ in Fig~2.1. \vadjust{\vskip 5pt\hbox 
to 0pt{\kern 10pt \includegraphics{3D_line.eps}\hss}\vskip -5pt}The point 
$A$ is a fixed point of this line, its radius-vector is $\bold r_0$. The point $X$ is a variable point, its radius-vector is $\bold r$. These two radius-vectors are related to 
each other by means of the equation
$$
\hskip -2em
\bold r=\bold r_0+\bold a\,t,
\mytag{2.1}
$$
where $\bold a$ is some non-zero vector on the line and $t$ is a scalar parameter. 
The equa\-lity \mythetag{2.1} is called the {\it vectorial parametric equation\/} 
of the line in the space (see \mycite{9}).\par
\parshape 5 180pt 180pt 180pt 180pt 180pt 180pt 180pt 180pt 0pt 360pt 
     The choice of the point $A$ on the line is not unique. Therefore the equation 
\mythetag{2.1} has some extent of ambiguity. In order to avoid this ambiguity 
\pagebreak non-parametric equations are used. Let's multiply both sides of the
equality \mythetag{2.1} by the vector $\bold a$ using the vector product\footnotemark\ 
operation. As a result we get
\footnotetext{\ It is also called the cross product, i\.\,e\. $[\bold x,\bold y]=\bold x\times
\bold y$.}
$$
\hskip -2em
[\bold r,\bold a]=[\bold r_0,\bold a].
\mytag{2.2}
$$
The product of two constant vectors in the right hand side of \mythetag{2.2} is a constant 
vector. If we denote it through $\bold b$, we get the equality
$$
\hskip -2em
[\bold r,\bold a]=\bold b\text{, \ where \ }\bold b\perp\bold a.
\mytag{2.3}
$$
The equality \mythetag{2.3} is known as the non-parametric {\it vectorial equation} of the 
line in the space (see \mycite{9}). Note that the vector $\bold b=[\bold r_0,\bold a]$ has
no ambiguity arising from the uncertainty in choosing the initial point $A$ on the line. Indeed, it is easy to see that $\bold b$ is invariant with respect to the transformation
$\bold r_0\to\bold r_0 + \bold a\,t$.\par
\head
3. The statement of the problem.
\endhead
     Let $X_1,\,\ldots,\,X_n$ be a group of points in the space given by their
radius-vectors $\bold r_1,\,\ldots,\,\bold r_n$. The linear regression problem
consists in finding a line given by the equation \mythetag{2.2} such that the
root mean square of the distances $d_1,\,\ldots,\,d_n$ from the points 
$X_1,\,\ldots,\,X_n$ to the line \mythetag{2.2} takes its minimal value:
$$
\hskip -2em
\bar d^{\kern 1.3pt 2}=\frac{1}{n}\sum^n_{i=1}d_i^{\kern 1.3pt 2}.
\mytag{3.1}
$$ 
\head
4. The solution of the problem.
\endhead
     The distance from the point $X_i$ to the line \mythetag{2.1} is given by the 
formula 
$$
\hskip -2em
d_i=\frac{|[\bold r_i-\bold r_0,\bold a]|}{|\bold a|}. 
\mytag{4.1}
$$
Without loss of generality we can assume that
$$
\hskip -2em
|\bold a|=1. 
\mytag{4.2}
$$
Then, taking into account $\bold b=[\bold r_0,\bold a]$ and \mythetag{4.2}, from
\mythetag{4.1} we derive
$$
\hskip -2em
d_i=|[\bold r_i,\bold a]-\bold b|. 
\mytag{4.3}
$$
Now we substitute \mythetag{4.3} into \mythetag{3.1}. As a result we obtain
$$
\hskip -2em
\bar d^{\kern 1.3pt 2}=|\bold b|^2 
-\frac{2}{n}\sum^n_{i=1}([\bold r_i,\bold a],\bold b)
+\frac{1}{n}\sum^n_{i=1}|[\bold r_i,\bold a]|^2. 
\mytag{4.4}
$$
The formula \mythetag{4.4} is an analog of the formula \thetag{2.3} in \mycite{4}.
The round brackets in \mythetag{4.4} denote the scalar product\footnotemark\ 
operation. 
\footnotetext{\ It is also called the dot product, i\.\,e\. $(\bold x,\bold y)=\bold x\cdot
\bold y$.}
\adjustfootnotemark{-2}
\mydefinition{4.1} A line given by the equation \mythetag{2.3} with
$|\bold a|=1$ is called an {\it optimal root mean square line\/} if
the quantity \mythetag{4.4} takes its minimal \pagebreak value.
\enddefinition
     The right hand side of \mythetag{4.4} is a quadratic polynomial with respect to
the components of the vector $\bold b$. It takes its minimal value if $\bold b$ is 
given by the formula 
$$
\hskip -2em
\bold b=\frac{1}{n}\sum^n_{i=1}[\bold r_i,\bold a].
\mytag{4.5}
$$
Substituting \mythetag{4.5} back into \mythetag{4.3}, we derive 
$$
\hskip -2em
\bar d^{\kern 1.3pt 2}=\frac{1}{n}\sum^n_{i=1}|[\bold r_i,\bold a]|^2
-\frac{1}{n^2}\sum^n_{i=1}\sum^n_{j=1}([\bold r_i,\bold a],[\bold r_j,\bold a]). 
\mytag{4.6}
$$
The formula \mythetag{4.6} is an analog of the formula \thetag{2.5} in \mycite{4}. 
Its right hand side is a quadratic form wit respect to the vector $\bold a$. We denote
it through $Q(\bold a,\bold a)$:
$$
\hskip -2em
Q(\bold a,\bold a)=\frac{1}{n}\sum^n_{i=1}|[\bold r_i,\bold a]|^2
-\frac{1}{n^2}\sum^n_{i=1}\sum^n_{j=1}([\bold r_i,\bold a],[\bold r_j,\bold a]) 
\mytag{4.7}
$$
and call the {\it non-linearity form} for a group of points in three-dimensional 
Euclidean space. Like the non-flatness form \thetag{2.14} in \mycite{4}, the 
non-linearity form \mythetag{4.7} is positive in the sense of the following inequality:
$$
Q(\bold a,\bold a)\geqslant 0\text{\ \ for \ }\bold a\neq 0. 
$$
Like in \mycite{4} one can draw some analogy to mechanics using the inertia tensor. 
However, we shall not do it now. We just note that like any quadratic form 
$Q(\bold a,\bold a)$ diagonalizes in some orthonormal basis associated with 
its primary axes.\par
     Let's introduce the following notation analogous to \thetag{2.6} in \mycite{4}:
$$
\hskip -2em
\bold r_{\text{cm}}=\frac{1}{n}\sum^n_{i=1}\bold r_i.
\mytag{4.8}
$$
The vector $\bold r_{\text{cm}}$ in \mythetag{4.8} is the radius-vector of the center 
of mass of a group of points $X_1,\,\ldots,\,X_n$ if assume that unit masses are placed
at each of these points. In terms of \mythetag{4.8} the formula \mythetag{4.5} is 
written as 
$$
\hskip -2em
\bold b=[\bold r_{\text{cm}},\bold a].
\mytag{4.9}
$$
Comparing \mythetag{4.9} with $\bold b=[\bold r_0,\bold a]$, we conclude that the optimal
line should pass through the center of mass of a group of points. Its direction is
determined by the non-linearity form $Q(\bold a,\bold a)$ according to the following 
theorem.
\mytheorem{4.1} A line is an optimal root mean square line
for a group of points if and only if it passes through the
center of mass of these points and if its direction vector $\bold a$
is directed along the primary axis of the non-linearity form $Q$ of
these points corresponding to its minimal eigenvalue.
\endproclaim
\head
5. Conclusion.
\endhead
    Theorem~\mythetheorem{4.1} solves the linear regression problem formulated in
Section 3. Its proof is obvious from the consideration preceding it. Practically
this theorem means that in order to find a line best fitting a group of points in 
three-dimensional Euclidean space one should find their center of mass and diagonalize the
symmetric matrix associated with their non-linearity form \mythetag{4.7}. In some 
cases this matrix can have two minimal eigenvalues $\lambda_1=\lambda_2<\lambda_3$. 
In these cases the shape of the group of points resembles a disc and hence there is
no preferable direction for the optimal line within the plane of this disc.\par
If $\lambda_1=\lambda_2=\lambda_3$, the shape of the group of points resembles a ball. 
In this case we have no preferable direction for the optimal line at all. 

\enddocument
\end